# Lone octupole and bulk magnetism in osmate 5d$^2$ double perovskites

S. W. Lovesey[1,2] and D. D. Khalyavin[1]

[1]ISIS Facility, STFC, Didcot, Oxfordshire OX11 0QX, UK

[2]Diamond Light Source Ltd, Didcot, Oxfordshire OX11 0DE, UK

**Abstract** Cubic double perovskites that host heavy ions with total angular momentum J = 2 can exhibit a singular magnetic state epitomized by a lone octupole and bulk ferro-type magnetism. It exists in the Chen - Balents Hamiltonian with a quadrupole interaction and competing exchange forces between the ions. Our symmetry inspired analysis mirrors the Dzyaloshinskii - Man'ko theory of latent antiferromagnetic ordering, and a 3-**k** collinear structure. Experimental tests for the singular state include neutron or resonant x-ray Bragg diffraction.

## I. INTRODUCTION

The double perovskite crystal structure (elpasolite) is composed of rock-salt ordered, corner-shared octahedra. Ions B and B′ in $A_2BB'O_6$ individually occupy face centred sublattices. In consequence, three-dimensional geometric magnetic frustration may occur when the B′ ion alone is magnetic. The structure can host heavy B′ ions, including W, Os and Re, with strong spin orbit coupling and concomitant entanglement of magnetic and lattice degrees of freedom. Structural, electronic and magnetic properties of materials $Ba_2BOsO_6$ with B = Mg, Zn, Ca, Cd have been measured recently in several exhaustive studies using a variety of experimental techniques [1 - 4]. In summary, no departure from cubic symmetry is observed, and thermodynamic anomalies consistent with a phase transition and muon spin rotation spectra indicate a low temperature (≈ 30 - 50 K) magnetic state with a very small Os magnetic dipole. Indeed, no magnetic Bragg spots have been identified in neutron diffraction patterns.

We predict magnetic properties of 5d$^2$ double perovskites using a symmetry informed analysis grounded on available experimental results. Our model uses magnetic space group $Pm\bar{3}m'$ in which magnetic dipoles are prohibited [5]. Instead, diffraction is caused by a solitary axial magnetic octupole that exists in the doubly-degenerate $\Gamma_3$ level (total angular momentum J = 2) with phase-shifted components that form a conjugate, or time reversed, pair of states [6, 7]. The octupole form factor in neutron diffraction h(κ) approaches zero for small Bragg angles, where the standard dipole form factor achieves its maximum. The maximum value of h(κ) occurs at a relatively large wave vector κ ≈ 5.68 Å$^{-1}$, and h(κ) achieves 57% of the maximum of the dipole form factor. Resonance enhancement of x-ray scattering at the osmium $L_1$ or K edges views the octupole through an electric quadrupole - electric quadrupole (E2-E2) absorption event.

A valuable parallel to our study is a much earlier theoretical investigation by Dzyaloshinskii and Man'ko [8]. The latent antiferromagnetic ordering discussed by the authors is the same as "latent" antiferromagnetic order of osmium octupoles in $Pm\bar{3}m'$. Notably, $Pm\bar{3}m'$ (with Os in 1$b$ and 3$d$ positions) yields precisely the ordering of octupoles that Dzyaloshinskii and Man'ko assign to one of the possibilities for magnetic dipoles in uranium dioxide (case III, page 917, with E < 0, and E < D). With interactions analogous to the usual exchange forces between dipoles and their local anisotropies, the thermodynamic theory carries over to octupole ordering and underpins it from a thermodynamic point of view. Specifically, Chen and Balents [7] represent interactions between $d^2$ ions on a face-centred cubic lattice by a quadrupole force and opposing exchange forces, and all three forces are anisotropic in position and spin space. The solitary magnetic octupole in $Pm\bar{3}m'$ belongs to one of seven mean-field states of the Hamiltonian (Section III.C.4 [7]). The centro-symmetric crystal class $m\bar{3}m'$ is not polar, and it is not compatible with conventional ferromagnetism involving dipole moments. While $m\bar{3}m'$ permits the piezomagnetic effect it is forbidden in the rock-salt structure $Fm\bar{3}m$ that belongs to the crystal class $m\bar{3}m$.

Available neutron and high-resolution x-ray diffraction patterns do not contain evidence of structural distortions violating F-centring, which is expected for $Pm\bar{3}m'$ [4]. This implies that experiments are not so sensitive to the magnetoelastic coupling. In turn, it might indicate that they are also not sensitive enough to see some other structural distortion, in particular deviation of the unit cell metric from the cubic one. It is not possible to calculate structure factors since the strength of the magnetoelastic coupling, and therefore the magnitude of the distortions, is totally unknown. Evidence of magnetic diffraction has been derived from the difference in neutron patterns taken at high and low temperatures, 280 and 4 K in the study of $Ba_2CaOsO_6$ by Thompson *et al*. [2]. Authors report difference data out to $\kappa = 3.58$ Å$^{-1}$ where the intensity of the predicted octupole is essentially half its maximum, with $[h(5.7)/h(3.58)]^2 = 1.97$. A more recent report of difference data, using 50 and 10 K, for three samples ($Ba_2BOsO_6$ with B = Zn, Mg, Ca) extends to $\kappa = 1.2$ Å$^{-1}$ [4, 9]. Neutron polarization analysis is the preferred technique for measuring the magnetic content of a Bragg spot indexed on the chemical structure.

Many materials possess magnetic octupoles formed with atomic states drawn from *nd* and *nf* atomic configuration, and known cases include both axial and polar varieties [10, 11]. Our candidate for the magnetic state of double perovskites hosting heavy ions with J = 2 is a singular case, however, in that magnetism is due to a lone octupole. Materials hosting *nd*-ions that possess axial or polar magnetic octupoles, in addition to multipoles of other rank, include NiO [12], $V_2O_3$ [13], $Cr_2O_3$ [14], $Sr_2IrO_4$ [15, 16], FeSe [17] and $Ca_3Ru_2O_7$ [18].

## II. MATERIAL PROPERTIES

Materials of interest retain the rock-salt structure with a lattice constant $a \sim 8.1$ Å to a very low temperature [1 - 4] using sites; Os (4$a$) (0, 0, 0), B (4$b$) (1/2, 1/2, 1/2) and A (8$c$) (1/4, 1/4, 1/4). Osmium ions occupy centrosymmetric sites with an ideal octahedral crystal field formed by six oxide ions (symmetry $m\bar{3}m$). Hexavalent osmium ($Os^{6+}$) has an incomplete $5d^2$ shell and a high-spin state $^3F$. In octahedral coordination, the ground state J = 2 is doubly degenerate ($\Gamma_3$ level) with magnetic projections M = 0, ±2 [4, 6, 7]. Projections obey $\Delta M$ = 0, ±2, ±4 because of a diad axis of rotation symmetry $2_z$. We use (ξ, η, ζ) with ξ = [1, −1, 0]/√2, η = [1, 1, −2] /√6, ζ = [1, 1, 1] /√3 for local Os coordinates, and reserve Cartesian (x, y, z) for axes parallel to cell edges that contain the tetrad symmetry axes.

## III. MULTIPOLES

Axial (parity-even) multipole of integer rank K are denoted $\langle T^K_Q \rangle$, where projections Q obey − K ≤ Q ≤ K, and angular brackets ⟨ ... ⟩ specify the time-average, or expectation value, of the enclosed spherical operator. The Hermitian property $\langle T^K_Q \rangle^* = (-1)^Q \langle T^K_{-Q} \rangle$ yields $\langle T^K_0 \rangle$ purely real. For x-ray Bragg diffraction enhanced by parity-even absorption event the time signature of $\langle T^K_Q \rangle$ depends on K alone, with K even (odd) charge-like (magnetic). All multipoles are time-odd for magnetic neutron scattering, of course. Multipoles engaged in magnetic neutron diffraction are defined in an Appendix; they have an odd rank when atomic states belong to a J-manifold, as in the present case [16].

## IV. CANDIDATE MAGNETIC STRUCTURES

Suitable magnetic structures are derived from $Fm\bar{3}m$. Axial magnetic multipoles alone are allowed with the specific constraint that dipoles are prohibited. A cubic magnetic motif described by a three-armed star with propagation vectors aligned along crystal axes $\mathbf{k}_1$ = (1, 0, 0), $\mathbf{k}_2$ = (0, 1, 0), $\mathbf{k}_3$ = (0, 0, 1) is required. Of three candidates retrieved from the conditional search one, $Pm\bar{3}m'$ (#221.95; [5]), permits investigation by magnetic diffraction of x-rays and neutrons, and corresponding diffraction amplitudes are the main subjects of this communication.

First, we describe the three candidates and argue that two are unacceptable, even though they meet the foregoing requirements, namely, cubic antiferromagnetism with no dipole moments. Osmium ions in $Pm\bar{3}m$ use non-equivalent sites (1a) and (3c) with site symmetries $m\bar{3}m$ ($O_h$) and 4/mm.m ($D_{4h}$), respectively. Site symmetry $m\bar{3}m$ restricts multipoles to two hexadecapoles $\langle T^4_0 \rangle$ and $\langle T^4_{+4} \rangle$ = √(5/14) $\langle T^4_0 \rangle$ for K ≤ 5. The triad axis of rotation symmetry $3_{xyz}$ is imposed in local coordinates (ξ, η, ζ) [19]. Charge-like hexadecapoles are engaged in resonant x-ray diffraction enhanced by an E2-E2 event [14], and even rank multipoles are absent in neutron diffraction by ions with an atomic state $\Gamma_3$ [16]. Multipoles do not satisfy the tetrad axis of rotation symmetry, $4_x$, in 4/mm.m. A second candidate, $Pn\bar{3}m$, uses sites (4c) with symmetry $\bar{3}m$ ($D_{3d}$) for osmium ions ($D_{3d}$ is appropriate for neptunium ions in $NpO_2$ using $Pn\bar{3}m$ [20]). All magnetic multipoles (K odd) are found to be purely real, whereas multipoles in $\Gamma_3$ are purely imaginary, as we see in Eqs. (2) and (5). The acceptable candidate, $Pm\bar{3}m'$, has osmium ions in sites (1b) and (3d) with symmetries $m\bar{3}m'$ and $4'/mm.m'$, respectively.

Of 10 possible magnetic multipoles, site symmetry $m\bar{3}m'$ forbids all but the axial octupole $\langle T^3_{+2} \rangle$. An argument for this remarkable result starts with the operation $2_z$ that imposes projections $Q = 2n$, with $n$ an integer. Symmetry $4_z'$ is satisfied by $n$ odd. A specific calculation reveals that symmetry $4_x'$ is satisfied by $K = 3$ alone, whereas $4_x$ cannot be satisfied by any K. The Hermitian property and the symmetry requirement $2_y \langle T^K_Q \rangle = (-1)^{K+Q} \langle T^K_{-Q} \rangle = \langle T^K_Q \rangle$ shows that $\langle T^3_{+2} \rangle$ is purely imaginary. Symmetries $2_z$, $2_y$, and $4_x'$ are among $4'/mm.m'$ operations. A proof that symmetry $2_{yz}'$ is satisfied uses the identity $2_{yz} = 2_y 4_x$. Evidently, $\langle T^K_0 \rangle = 0$ for K odd for both sites, and $K = 1$ or $3$ for a $J = 2$ manifold of states. This result and $Q = 2n$ with $n$ odd prohibit a dipole. Fig. 1 depicts the arrangement of magnetic octupoles in $Pm\bar{3}m'$.

## V. NEUTRON DIFFRACTION

A general form of a partner in $\Gamma_3$ is,

$$|g\rangle = \alpha|0\rangle + \beta[|+2\rangle + |-2\rangle], \tag{1}$$

where $|M\rangle = |J = 2, M\rangle$. The time-reversed (conjugate) state of $\gamma|j, m\rangle = \gamma^* (-1)^{j-m} |j, -m\rangle$, where $\gamma$ is a c-number. In consequence, $\Gamma_3$ and $\Gamma_5$ states of the $J = 2$ manifold are even (non-magnetic) and odd (magnetic), respectively (alternatively, components of the doublet $\Gamma_3$ can be chosen purely real and magnetism is forbidden). In Eq. (1) the coefficients are $\alpha = (\alpha' + i\alpha'')$ with $\alpha'$, $\alpha''$, and $\beta$ purely real, and $\{|\alpha|^2 + 2\beta^2\} = 1$ for normalization. Notably, the partner (1) is contained in the mean-field state Eq. (52) in Ref. [7]. Partners in the $\Gamma_3$ conjugate pair are separated in energy by a molecular field, and $|g\rangle$ is chosen as the ground-state. Evidently, a magnetic dipole is prohibited and $\langle T^K_0 \rangle = 0$ for K odd, as required by site symmetry. On the other hand,

$$\langle g|T^3_{+2}|g\rangle = \langle T^3_{+2} \rangle = - i \, (12/7) \sqrt{(1/35)} \, \alpha'' \beta \, h(\kappa), \tag{2}$$

for the saturation value of the octupole in $Pm\bar{3}m'$. The form factor $h(\kappa) = [\langle j_2(\kappa) \rangle + (10/3) \langle j_4(\kappa) \rangle]$ is displayed in Fig. 2, and $\kappa = \{(4\pi/\lambda) \sin(\theta)\}$ where $\theta$ is the Bragg angle, as in Fig. 3, and $\lambda$ the neutron wavelength. Radial integrals $\langle j_m(\kappa) \rangle$ are calculated from results in Ref. [21]. The octupole form factor is a maximum for $\kappa \approx 5.68$ Å$^{-1}$ where $h(\kappa) \approx 0.567$.

Miller indices for the parent structure $Fm\bar{3}m$ satisfy F-centring with $H_o + L_o$, $K_o + L_o$, $H_o + K_o$ simultaneously even. The basis vectors of $Pm\bar{3}m'$ are $\{(0, -1, 0), (-1, 0, 0), (0, 0, -1)\}$ with Miller indices $h = - K_o$, $k = - H_o$, $l = - L_o$. There are no systematic absences in the diffraction pattern that arise from translational components of the symmetry elements alone. Osmium sites are (1$b$) at (1/2, 1/2, 1/2), and (3$d$) at (1/2, 0, 0), (0, 0, 1/2), (0, 1/2, 0) with environments at the second and third sites related to (1/2, 0, 0) by tetrad operations $4_y'$ and $4_z'$, respectively. Diffraction amplitudes are constructed from an electronic structure factor $\Psi^K_Q = [\exp(i\boldsymbol{\kappa} \cdot \mathbf{d}) \langle T^K_Q \rangle_\mathbf{d}]$, where the Bragg wavevector $\boldsymbol{\kappa} = (h, k, l)$ and the implied sum is over all osmium sites. For the singular case in hand, $K = 3$. In general, a bulk (macroscopic) property associated with $\langle T^K_Q \rangle$ is permitted if one component of $\Psi^K_Q$ is different from zero for $\kappa = 0$, and this quantity is proscribed by symmetry of the magnetic crystal class.

A theory of magnetic neutron scattering used in subsequent calculations is outlined in an Appendix [22, 23]. A unit vector $(p, q, r) = (h, k, l) [h^2 + k^2 + l^2]^{-1/2}$ proves useful in expressions for the amplitudes of magnetic diffraction. The latter are,

$$\langle Q_{\perp,x} \rangle = C\,(q\,r)\,(1 - 3\,p^2),\ \langle Q_{\perp,y} \rangle = C\,(p\,r)\,(1 - 3\,q^2),$$

$$\langle Q_{\perp,z} \rangle = C\,(p\,q)\,(1 - 3\,r^2), \tag{3}$$

with a common spatial phase $[(-1)^h + (-1)^k + (-1)^l]$ for sites (3d). The common spatial phase $= [(-1)^{h+k+l}]$ for (1b). Both phases are different from zero for the trivial wave vector $h = k = l = 0$. In consequence, bulk octupole order is allowed in $Pm\bar{3}m'$. Conventional ferromagnetic order using dipoles does not exist, since $\langle \mathbf{T}^1 \rangle = 0$, which accords with properties of the magnetic crystal class. The purely real pre-factor $C = \{(1/2)\,\sqrt{(105/2)}\,\langle T^3_{+2} \rangle''\}$ with the result (2) for $\langle T^3_{+2} \rangle'' \propto (xyz)$ derived for $\Gamma_3$.

Intensities of magnetic Bragg spots $|\langle \mathbf{Q}_\perp \rangle|^2$ can be different from zero for all reflections. Intensities for independent sites add, i.e., intensity of a magnetic Bragg spot $= \{|\langle \mathbf{Q}_\perp(1b) \rangle|^2 + |\langle \mathbf{Q}_\perp(3d) \rangle|^2\}$ for $Pm\bar{3}m'$. It follows from (3) that $|\langle \mathbf{Q}_\perp \rangle|^2$ is a function of $p^2$, $q^2$ and $r^2$, which is explicit in the useful result (A8). Conditions for optimal intensity can be estimated by treating p, q, r as continuous variables. Using (A8), we find $q = 0$, $p = r = 1/\sqrt{2}$, leading to an optimal magnetic neutron scattering intensity $= [10\,C^2\,(1/4)]$, where the factor 10 is from spatial phases. The established p, q, r, together with $\kappa \approx 5.68$ Å$^{-1}$ for the position of the maximum of the octupole form factor in $\langle T^3_{+2} \rangle''$ reveals that intensity is a maximum at $\kappa = (2\pi/a)\,(0, 6, 6)$. The powder average intensity (A5) amounts to 18% of $[10\,C^2\,(1/4)]$.

Magnetic signal atop relatively strong nuclear scattering can be disentangled by polarization analysis [24]. We report some values of the spin-flip intensity,

$$SF = |\langle \mathbf{Q}_\perp \rangle - \mathbf{P}\,(\mathbf{P} \bullet \langle \mathbf{Q}_\perp \rangle)|^2, \tag{4}$$

where $\mathbf{P}$ is a *unit* vector in the direction of the polarization in the primary neutron beam. For $\mathbf{P}$ parallel to the Bragg wavevector $\kappa \propto (p, q, r)$ the result is $SF = |\langle \mathbf{Q}_\perp \rangle|^2$. Using (p, p, r) as an example, we consider the opposite extreme of $\mathbf{P} \perp \kappa$. First, $\mathbf{P} = (r, r, -2p)/\sqrt{2}$ yields $\mathbf{P} \bullet \langle \mathbf{Q}_\perp \rangle = [p\,\sqrt{2}\,(r^2 - p^2)]$ and $SF = 0$. On the other hand, $\mathbf{P} = (-1, 1, 0)/\sqrt{2}$ leads to $\mathbf{P} \bullet \langle \mathbf{Q}_\perp \rangle = 0$ and $SF = [20\,C^2\,p^2\,(1 - 3p^2)^2]$ is readily obtained from (A8).

## VI. RESONANT X-RAY DIFFRACTION

Resonant x-ray Bragg diffraction enhanced by an electric quadrupole - electric quadrupole (E2-E2) event accesses magnetic multipoles with rank K = 1 and 3 (an E1-E1 event does not access K = 3 [27]). Corresponding multipoles, denoted $\langle t^K_Q \rangle$, depend on the quantum numbers that define the absorption event [14]. In the context of $Pm\bar{3}m'$ we need consider K = 3 alone. Multipoles observed through absorption at the Os K-edge ($E_K \approx$ 73.871 keV, 1s core state) and $L_1$-edge ($E_L \approx$ 12.968 keV, 2s core state) depend on the orbital state and not the spin state of the resonant ion [12]. To the extent that radial wave functions in a highly ionized ion are hydrogenic in form, $\langle ns|R^2|5d \rangle$ is proportional to $(1/Z_c)^2$ where $Z_c$ is the effective nuclear core charge seen by the jumping electron. And an explicit calculation for relative intensities at the two edges yields $\{[\langle 2s|R^2|5d \rangle E_L]/[\langle 1s|R^2|5d \rangle E_K]\}^2 = 0.32$ [19]. The saturation value of the Os octupole,

$$\langle t^3_{+2} \rangle = - i\, (3/7)\, \sqrt{(2/7)}\, \alpha''\, \beta, \qquad (5)$$

is purely imaginary, like its counterpart (2) in neutron diffraction.

For the purpose of immediate illustration, we consider diffraction in the rotated channel of polarization $\pi'\sigma$, with polarization states depicted in Fig. 3, and examine two Bragg wave vectors $\kappa \propto (1, 0, 0)$ and $\kappa \propto (1, 1, 0)$. The crystal can be rotated about $\kappa$ by an angle $\psi$, and crystal axes a & b are contained in the plane of scattering at the origin of an azimuthal-angle scan $\psi = 0$. A diffraction amplitude $F(\pi'\sigma)$ is easily calculated from general expressions for all four channels of polarization [25]. First, the Bragg wave vector p = 1, q = r = 0:

$$F(\pi'\sigma) = -\,(1/8)\,\sqrt{3}\,\langle t^3_{+2} \rangle\,[(-1)^h + (-1)^k + (-1)^l]\,[\sin(\theta) + \sin(3\theta)]\,\sin(2\psi). \qquad (6)$$

The spatial phase factor is correct for sites (3d), and it is replaced by $[(-1)^{h+k+l}]$ for (1b). The amplitude has simple two-fold rotation symmetry since $\kappa$ is parallel to a high-symmetry axis. As a second example, $p = q = 1/\sqrt{2}, r = 0$:

$$F(\pi'\sigma) = -\,(1/4)\,\sqrt{3}\,\langle t^3_{+2} \rangle\,\cos(\theta)\,\sin(\psi)\,[(-1)^h + (-1)^k + (-1)^l]$$

$$\times\,[\cos^2(\theta) + \{3\cos^2(\theta) - 2\}\cos(2\psi)]. \qquad (7)$$

Amplitudes $F(\pi'\sigma)$ for the two wave vectors have very different dependences on $\psi$, apart from the fact that (6) and (7) are zero at the azimuthal origin.

## VII. CONCLUSIONS AND DISCUSSION

In summary, we propose that osmate double perovskites, e.g., $Ba_2CaOsO_6$, possess an unusual magnetic state that supports a macroscopic ferro-octupole moment (more precisely ferri-octupole moment). It is compatible with entropy release, zero-field oscillations in μSR measurements, and the absence of magnetic Bragg spots in a conventional diffraction pattern [1 - 4]. The magnetic state in question is defined by the cubic space group $P m\bar{3}m'$ with Os ions in two independent sites, both endowed with symmetry that admits a lone octupole. Additional notable features include, structural distortions that violate F-centring in the parent rock-salt structure, permitted by magneto-elastic couplings in $Pm\bar{3}m'$, a piezomagnetic effect, and a one-to-one correspondence with the Dzyaloshinskii - Man'ko theory of latent antiferromagnetic ordering [8]. Magnetic diffraction of neutrons and x-rays are discussed in Sections V and VI. In the case of neutron diffraction, the octupole form factor, displayed in Fig. 2 for $Os^{6+}$, specifies relatively large wave vectors not used hitherto.

It is appropriate to mention an interpretation of an uncommon electronic phase transition in neptunium dioxide at a temperature $\approx 25.5$ K that uses reduction in spatial symmetry from $Fm\bar{3}m$ (#225) to $Pn\bar{3}m$ (#224) [26]. Neptunium site symmetry descends from $O_h$ ($m\bar{3}m$), with all multipoles prohibited other than $\langle T^4 \rangle$, to $D_{3d}$ ($\bar{3}m$) that allows multipoles with even and odd ranks $K \geq 2$. The two structures, $Fm\bar{3}m$ and $Pn\bar{3}m$, possess identical extinction rules yielding identical conventional Bragg diffraction patterns, although extinction rules arise from simple translations in the former and screw axes and glide planes in the latter. The interpretation is founded on an observation of diffraction by charge-like quadrupoles in resonance enhanced Bragg diffraction, so-called Templeton-Templeton scattering that occurs in basis - forbidden Bragg spots [20, 26]. In contrast, our proposal for cubic double perovskites uses a magnetic transition $Fm\bar{3}m \rightarrow Pm\bar{3}m'$ that results in an absence of extinction rules, since $Pm\bar{3}m$ is one of the 73 symmorphic space-groups (no screw axes parallel to principal directions and no glide planes perpendicular to the principal directions).

**Acknowledgements** Dr V. Scagnoli drew our attention to Ref. [4]. SWL is grateful to Dr K. S. Knight for ongoing guidance on Bragg diffraction, and perusal of our communication in its making. Professor S. P. Collins advised on the prospect of x-ray Bragg diffraction enhanced by an osmium E2-E2 absorption event.

## APPENDIX

We outline a theory for the magnetic scattering of neutrons by unpaired electrons in an atomic shell. As early as (1953), Trammell calculated the amplitude in the context of diffraction by rare earth ions [28]. He did not introduce multipoles with discrete symmetries, and his final result is unnecessarily complicated with different radial integrals for the spin and orbital components of the interaction operator (A1) [29, 30].

The magnetic scattering operator $\mathbf{Q}_\perp = \{\kappa^{-2} [\boldsymbol{\kappa} \times (\mathbf{Q} \times \boldsymbol{\kappa})]\}$ with an intermediate operator,

$$\mathbf{Q} = \exp(i\mathbf{R}_j \cdot \boldsymbol{\kappa}) [\mathbf{s}_j - \kappa^{-2} (i/\hbar)(\boldsymbol{\kappa} \times \mathbf{p}_j)], \tag{A1}$$

and the implied sum is over all unpaired electrons. In (A1), **R** and **p** are conjugate operators for electronic position and linear momentum, respectively. It can be shown that, [14, 22, 23]

$$\mathbf{Q} = \sum_{K'} [(2K' + 1)/(K' + 1)] (2K' - 1)^{1/2} \{\mathbf{C}^{K'-1}(\boldsymbol{\kappa}) \otimes \mathbf{T}^{K'}\}^{K'}$$

$$+ i \sum_{K} (2K + 1)^{1/2} \{\mathbf{C}^{K}(\boldsymbol{\kappa}) \otimes \mathbf{T}^{K}\}^{K}, \quad (A2)$$

where $\mathbf{C}^a(\boldsymbol{\kappa})$ is a spherical harmonic normalized such that $C^1(\boldsymbol{\kappa}) = \boldsymbol{\kappa}$. (With J an integer, $\langle \mathbf{T}^K \rangle$ and $\mathbf{C}^K(\mathbf{R})$ are equivalent under simple rotations, and a factor i in $\mathbf{C}^K(\mathbf{R})$ is the time signature. Specifically, $\langle T^3_{+2} \rangle'' \propto (xyz)$ from $C^3_{+2}(\mathbf{R}) \propto [(x + iy)^2 z]$.) A tensor product $\{\mathbf{C}^a(\boldsymbol{\kappa}) \otimes \mathbf{T}^b\}^c$ is defined by,

$$\{\mathbf{C}^a(\boldsymbol{\kappa}) \otimes \mathbf{T}^b\}^c{}_\gamma = \sum_{\alpha,\beta} C^a{}_\alpha(\boldsymbol{\kappa}) T^b{}_\beta (a\alpha\, b\beta \,|\, c\gamma). \quad (A3)$$

The Clebsch-Gordan coefficient in (A3) is purely real, and related to the standard Wigner 3-j symbol,

$$(a\alpha\, b\beta \,|\, c\gamma) = (-1)^{-a+b-\gamma} \sqrt{(2c + 1)} \begin{pmatrix} a & b & c \\ \alpha & \beta & -\gamma \end{pmatrix}. \quad (A4)$$

Allowed multipoles in (A2) are K even (= 2, ..., 2$l$) and K' odd (= 1, 3, ..., 2$l$ + 1), where $l$ is the orbital angular momentum of the shell. Multipoles with K even are forbidden in a J-manifold and they are absent in calculations using the osmium $\Gamma_3$ ground-state.

An average of the neutron cross-section with respect to directions of the scattering wavevector yields a powder diffraction pattern,

$$I = (1/4\pi) \int d\hat{\boldsymbol{\kappa}} \, \{\langle \mathbf{Q}_\perp \rangle \cdot \langle \mathbf{Q}_\perp \rangle\} = \sum_{K,Q} [3/(2K + 1)] |\langle T^K{}_Q \rangle|^2 + \sum_{K',Q'} [3/(K' + 1)] |\langle T^{K'}{}_{Q'} \rangle|^2, \quad (A5)$$

with K even and K' odd as in (A2).

An approximate dipole $\mathbf{T}^1 \approx (1/3) \{2\mathbf{S} \langle j_0(\kappa) \rangle + \mathbf{L} [\langle j_0(\kappa) \rangle + \langle j_2(\kappa) \rangle]\}$ is often used in the interpretation of elastic and inelastic scattering. The radial integral $\langle j_0(0) \rangle = 1$, while $\langle j_m(0) \rangle = 0$ for m ≥ 2. In the general case, reduced matrix-elements for the two multipole operators in (A2) are,

$$(\theta\|\mathbf{T}^{K'}\|\theta') = -(-1)^{J'-J} (2J + 1)^{1/2} \{A(K' - 1, K') + B(K' - 1, K')\}, \quad (A6)$$

$$(\theta\|\mathbf{T}^K\|\theta') = -i(-1)^{J'-J} (2J + 1)^{1/2} B(K, K), \quad (A7)$$

where purely real A(K' − 1, K'), B(K' − 1, K') and B(K, K) are tabulated for $d^n$ and $f^n$ configurations [23]. Contributions A(K'', K') and B(K'', K') are created by the orbital and spin parts of (A1), respectively. The shorthand for quantum labels is θ = J, S, L and θ' = J', S', L' (Russell-Saunders coupling scheme). In a J-manifold drawn from an atomic *nd*-shell K' = 1, 3 and 5, with K′ = 5 proportional to B(4, 5) alone. However, K′ = 5 is absent in the first half of the $nd$-shell since 2J < 5 for these ions.

Using results (3) for the components of $\langle \mathbf{Q}_\perp \rangle$,

$$|\langle \mathbf{Q}_\perp \rangle|^2 = (C/2)^2 \left[ \sin^2(2\chi) + \{\sin^2(\chi) \sin(2\varphi)\}^2 \{9 \sin^2(\chi) - 8\} \right], \tag{A8}$$

where $(p, q, r) = (\sin(\chi) \cos(\varphi), \sin(\chi) \sin(\varphi), \cos(\chi))$.

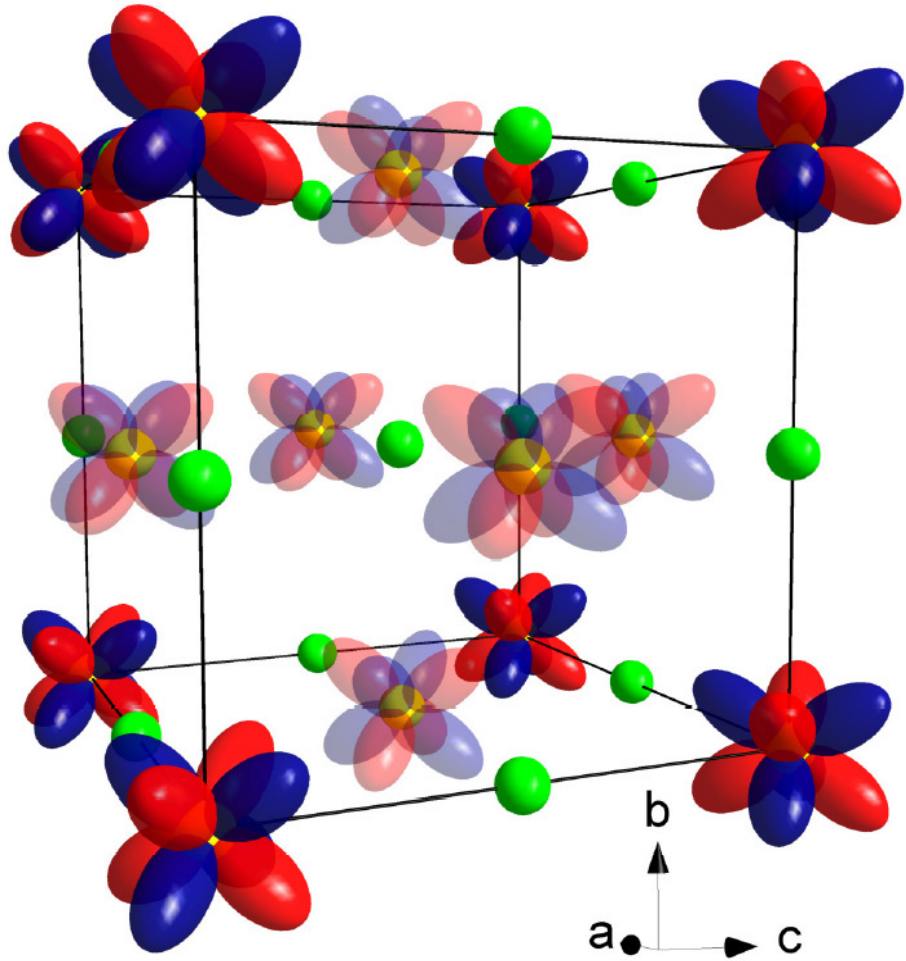

Fig. 1. Magnetic octupoles $\langle T^3_{+2}\rangle'' \propto$ (xyz) in $Pm\bar{3}m'$. Os octupoles in two independent crystallographic positions are shown with different degree of transparency (less transparent in 1$b$ and more transparent in 3$d$ position). B-cations are shown as green spheres. The structure is shown in the settings of the parent paramagnetic $Fm\bar{3}m$ space group and Ba and O are omitted for clarity.

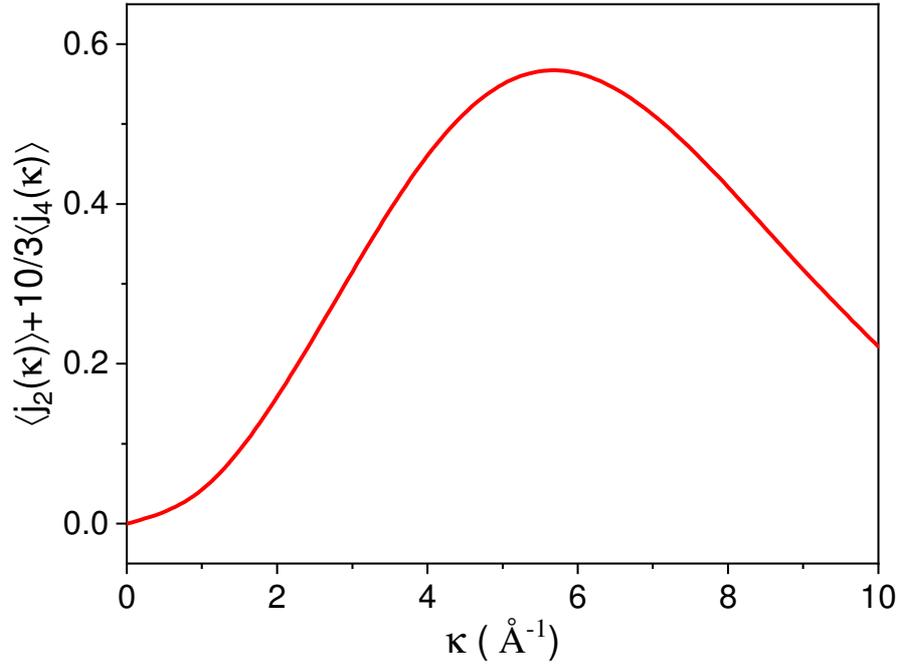

Fig. 2. Octupole form factor $h(\kappa) = \{\langle j_2(\kappa)\rangle + (10/3)\langle j_4(\kappa)\rangle\}$ for $Os^{6+}$, $5d^2$. Radial integrals $\langle j_m(\kappa)\rangle$ are derived from results published in Ref. [21]. Maximum value occurs at the wavevector $\kappa \approx 5.68$ Å$^{-1}$ with $h(\kappa) \approx 0.567$. A dipole contribution to scattering $\langle \mathbf{T}^1\rangle$ is proportional to $\langle j_0(\kappa)\rangle$ that is unity in the forward direction, by definition, and it decreases monotonically to zero at $\kappa \approx 5.0$ Å$^{-1}$. Magnetic dipoles are forbidden in $Pm\bar{3}m'$, however.

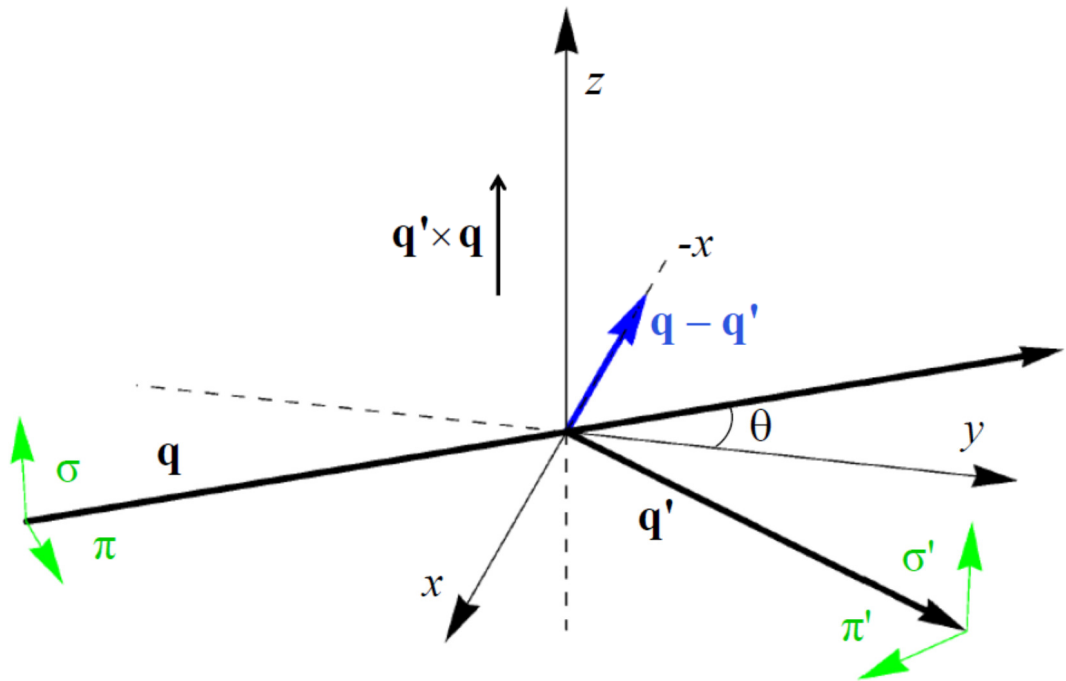

Fig. 3. X-ray diffraction. Primary (σ, π) and secondary (σ′, π′) states of polarization. Corresponding wave vectors **q** and **q**′ subtend an angle $2\theta$. Cell edges of $A_2BB'O_6$ and depicted Cartesian co-ordinates (x, y, z) coincide in the nominal setting of the crystal.